\documentclass{article}
\usepackage{LaThuileFPSpro}
\usepackage{epsfig}

\begin{document}
\title{ 
  RECENT RESULTS FROM CLEO-c
  }
\author{
  Alex Smith        \\
  {\em University of Minnesota}\\
  {\em 116 Church St. S.E.}\\
  {\em Minneapolis, MN Minnesota 55455}\\
  }
\maketitle

\baselineskip=11.6pt

\begin{abstract}
  This paper describes recent preliminary results from the CLEO-c
  experiment using an initial $\sim 60\ {\rm  pb}^{-1}$ sample of data
  collected in $e^{+}e^{-}$ collisions at a center of mass energy
  around the mass of 
  the $\psi(3770)$.  A first measurement of the
  branching fraction ${\cal B}(D^{+}\rightarrow
  \mu^{+}\nu) = (3.5\pm 1.4\pm 0.6)\times 10^{-4}$ and the
  corresponding decay constant $f_{D} = (202\pm 41 \pm 17)$~MeV has
  been made.  Several charged and neutral $D$ meson absolute exclusive
  semileptonic branching fractions have been measured, including first
  measurements of the branching fractions ${\cal B}(D^{0}\rightarrow
  \rho^{-}e^{+}\nu) = (0.19 \pm 0.04 \pm 0.02)\%$ and 
  ${\cal B}(D^{+}\rightarrow \omega
  e^{+}\nu)=(0.17 \pm 0.006 \pm 0.01)\%$.  
  Estimated uncertainties for inclusive $D$ semileptonic
  decay modes are also presented.  Fits to single and double $D$
  tagged events are used to extract absolute branching fractions of
  several hadronic $D$ decay modes and $D\overline{D}$ production
  cross sections.  Most of these results from this small preliminary
  sample are already of greater sensitivity than previously
  published results.

\end{abstract}
\newpage
\section{Introduction}
The CLEO-c physics program is focused on the study of charm decays in
$e^{+}e^{-}$ collisions in the CESR-c storage ring at
energies near the $\psi(3770)$ and $J/\psi$ resonances and above 
$D_{s}\overline{D_{s}}$ threshold.  The results presented in this
paper are based on approximately 60~pb$^{-1}$ of data collected at the
$\psi(3770)$, just above $D\overline{D}$ threshold.  

For many electroweak quantities measured by the $B$ factories at SLAC
and KEK, in particular many that 
contribute to constraining the CKM unitarity triangle\cite{CKM}, the
precision is 
limited by theoretical uncertainties rather than experimental precision.
One of the primary goals of these measurements is the calibration and
validation of lattice QCD.  Lattice QCD will soon be able to
predict many quantities such as the decay constants $f_{D}$ and
$f_{D_{S}}$ of $D$ and $D_{S}$ mesons with few percent uncertainties.
Measurement of $f_{D}$ will lead to a determination of $f_{B}$, since
lattice QCD can predict the ratio $f_{B}/f_{D}$ better than the
absolute decay constants.
It is critical, however, that the uncertainties of the lattice
calculations be 
verified by experimental measurements.  CLEO-c measurements of
absolute branching fractions and 
form factors for a full isospin set of semileptonic decays will
provide a stringent test of form factor calculations and models.

In addition to verification of lattice QCD, CLEO-c will improve on the
existing measurements of $|V_{cs}|$ and $|V_{cd}|$ and measure absolute
branching fractions for many important hadronic normalization modes which
contribute significant uncertainties to important measurements 
at higher energies.

\section{Purely Leptonic D Meson Decay Absolute Branching Fraction and the Decay Constant $f_{D}$}

The decay constant $f_{D}$ is an important parameter which quantifies
the annihilation probability of the valence quarks of the $D$ meson. 
This parameter can be determined from the absolute branching fraction
${\cal B}(D^{+}\rightarrow \mu^{+}\nu)$.  A first measurement of the
absolute branching fraction of the decay $D^{+}\rightarrow \mu^{+}\nu$ 
was recently made by CLEO-c.   

The analysis relies on fully reconstructing or ``tagging'' one $D$ or
$\overline{D}$ meson in the $D\overline{D}$ pair produced in the
$\psi(3770)$ decay.  This technique works quite well at the
$\psi(3770)$ resonance, since there is not enough energy in the event
to produce hadrons other than the $D\overline{D}$ pair.  Using the
decay modes $D^{-}\rightarrow K^{+}\pi^{-}\pi^{-},\
K^{+}\pi^{-}\pi^{-}\pi^{0},\ \overline{K_{S}^{0}}\pi^{-},\
\overline{K_{S}^{0}}\pi^{-}\pi^{-}\pi^{+},\
\overline{K_{S}^{0}}\pi^{-}\pi^{0}$ for the tag side $D$, an
efficiency of approximately 25\% for the tag reconstruction is achieved.

\begin{figure}[htp]
  \centerline{
    \epsfxsize=3.00in
    \epsffile{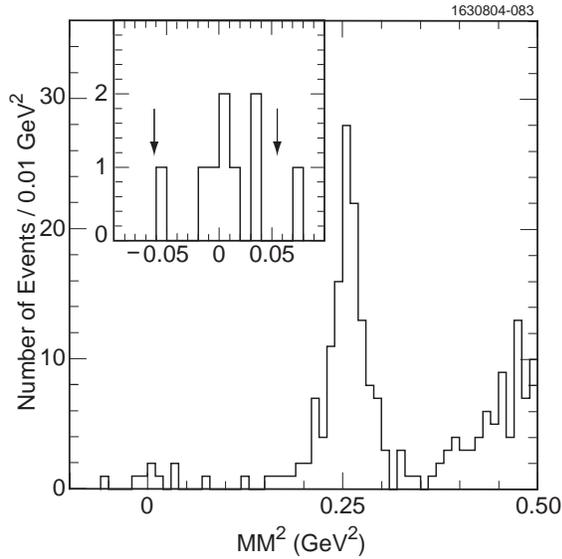}
  }
  \caption{\it Missing mass distribution of $D^{+}\rightarrow \mu^{+}\nu$
  candidates.  The signal peaks at zero missing mass.}
  \label{fig:missmass}
\end{figure}

A very pure sample of $28651 \pm 207$ tagged $D$ mesons is selected.
The tag $D$ meson is combined with an additional charged track of the
correct sign, presumed to be a muon.  The distribution of 
``missing mass squared'',
defined to be $M_{\rm miss}^{2}\equiv (E_{\rm beam} - E_{\mu^{+}})^{2} -
(\vec{p}_{D^{-}} - \vec{p}_{\mu^{+}})^{2}$, is shown in
Fig.~\ref{fig:missmass}.  A significant and well-defined peak of
eight events at the neutrino mass around zero is observed.  The peak
at 0.25~GeV$^{2}/c^{4}$ 
corresponds to a background from decays to the $K^{0}_{L}\pi^{+}$
final state, which is well separated from the signal region.  The
total contribution of backgrounds in the signal region is estimated in
a maximum likelihood fit to
be one event.  This leads to a branching fraction of ${\cal B}(D^{+}
\rightarrow \mu^{+}\nu) = (3.5 \pm 1.4 \pm 0.6)\times 10^{-4}$ and a
$D$ meson decay constant of $f_{D} = (202 \pm 41 \pm 17)$~MeV, where
the first uncertainty is statistical and the second is 
systematic\cite{DMuNuPRD}.

\section{Absolute Branching Fractions of Exclusive Semileptonic D
  Meson Decays }

\begin{figure}[htp]
  \centerline{
    {\large a)}
    \epsfxsize 5.7cm
    \epsffile{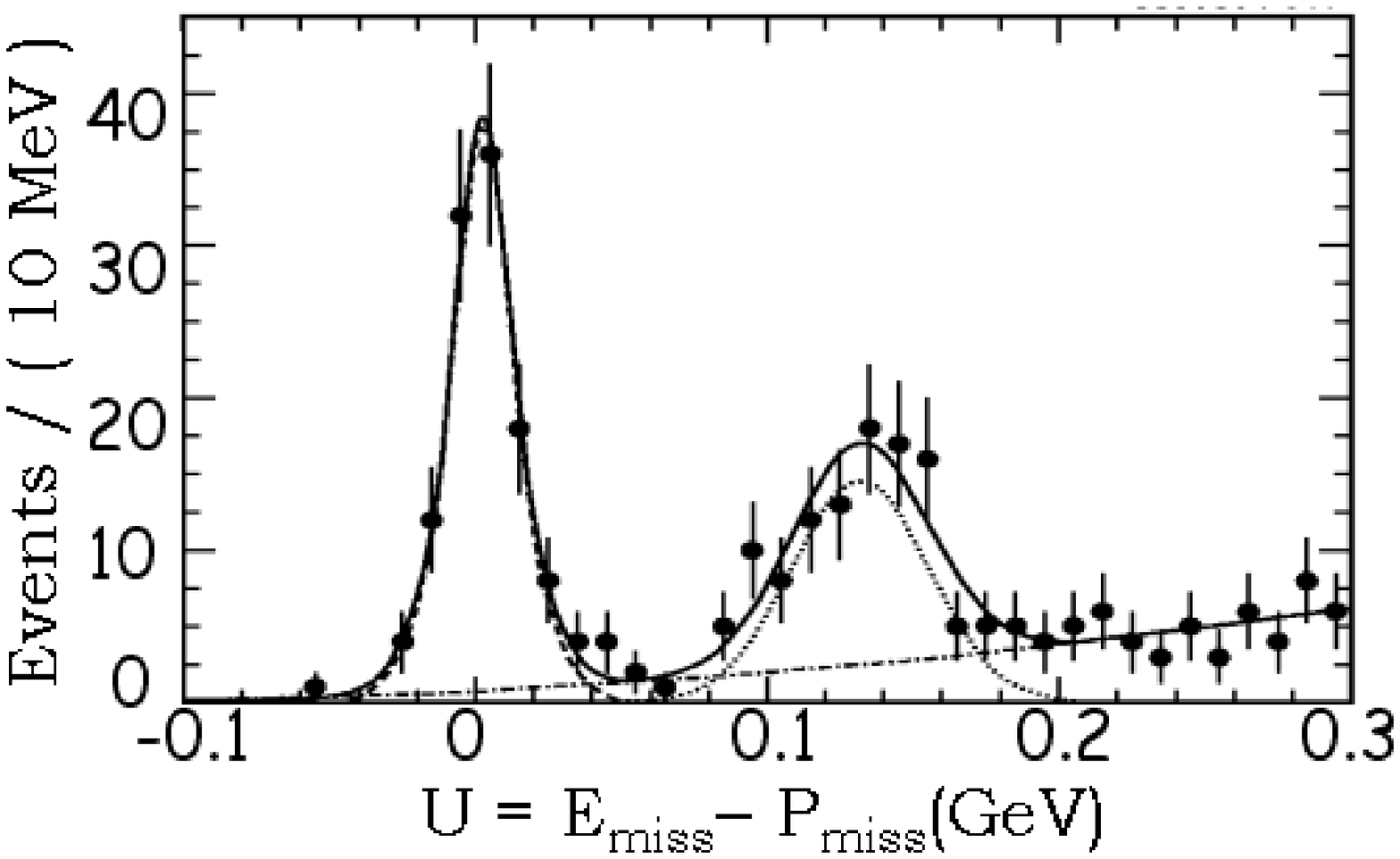}
    {\large b)}
    \epsfxsize=5.7cm
    \epsffile{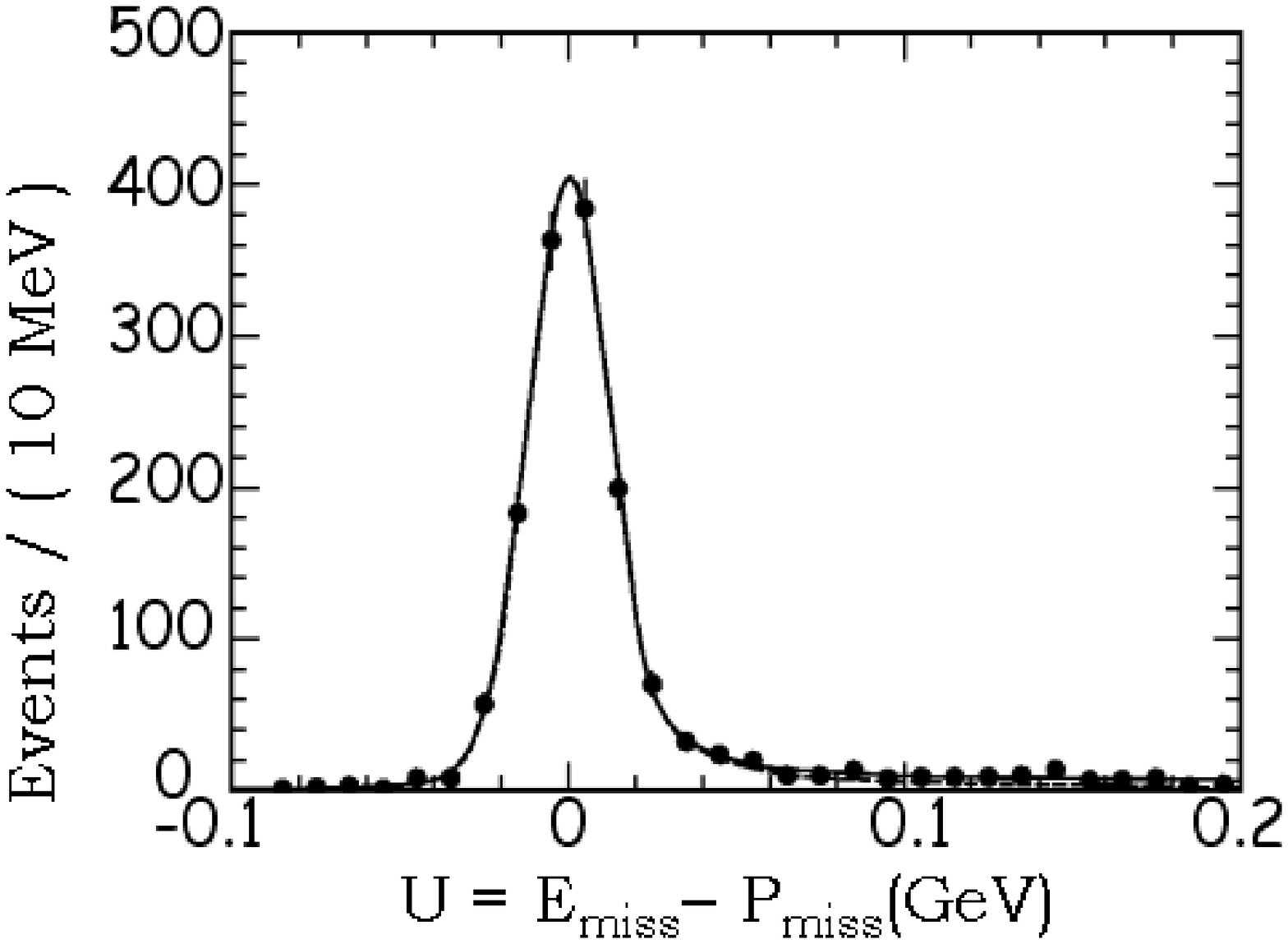}
  }
  \centerline{
    {\large c)}
    \epsfxsize=5.7cm
    \epsffile{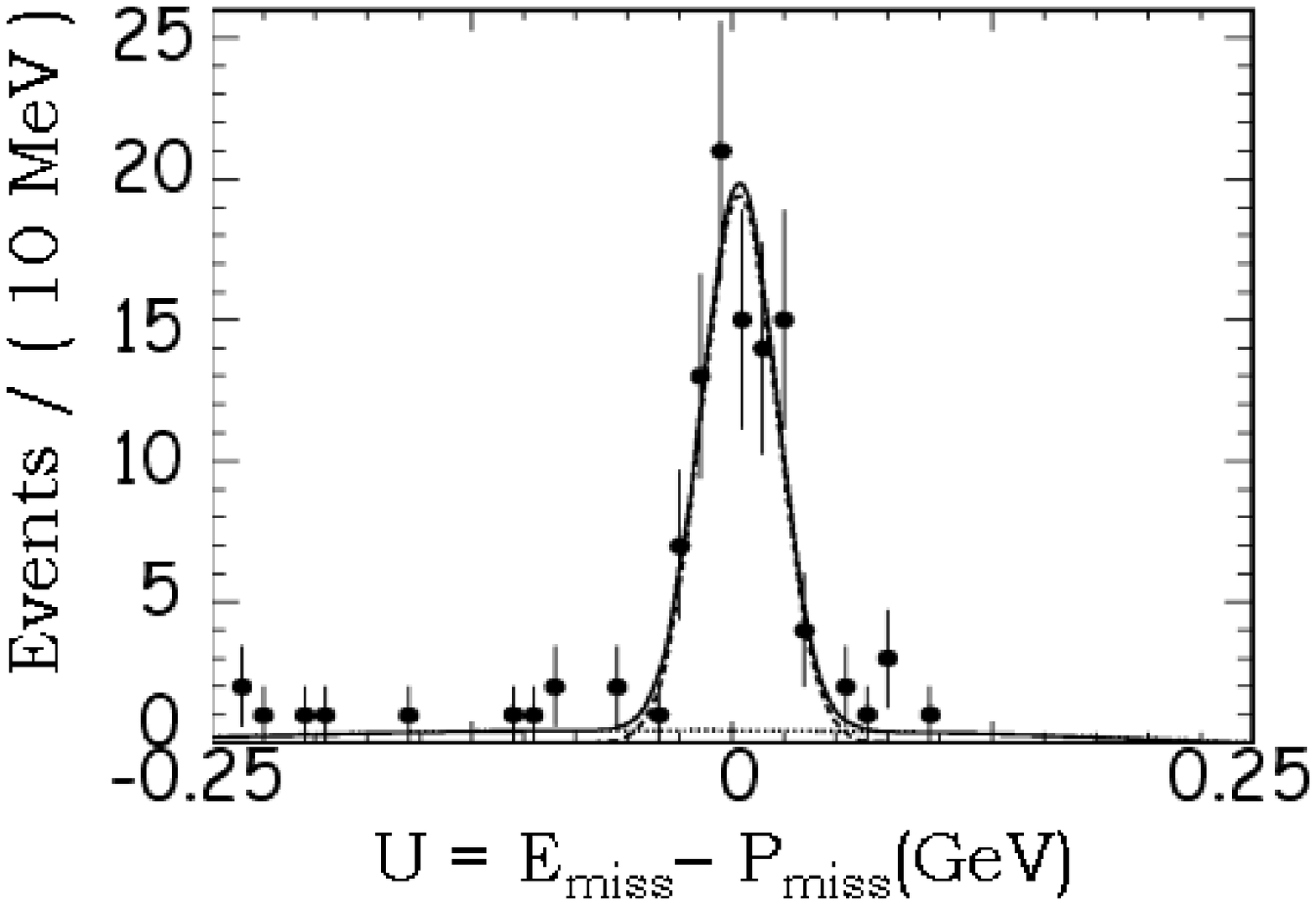}
    {\large d)}
    \epsfxsize=5.7cm
    \epsffile{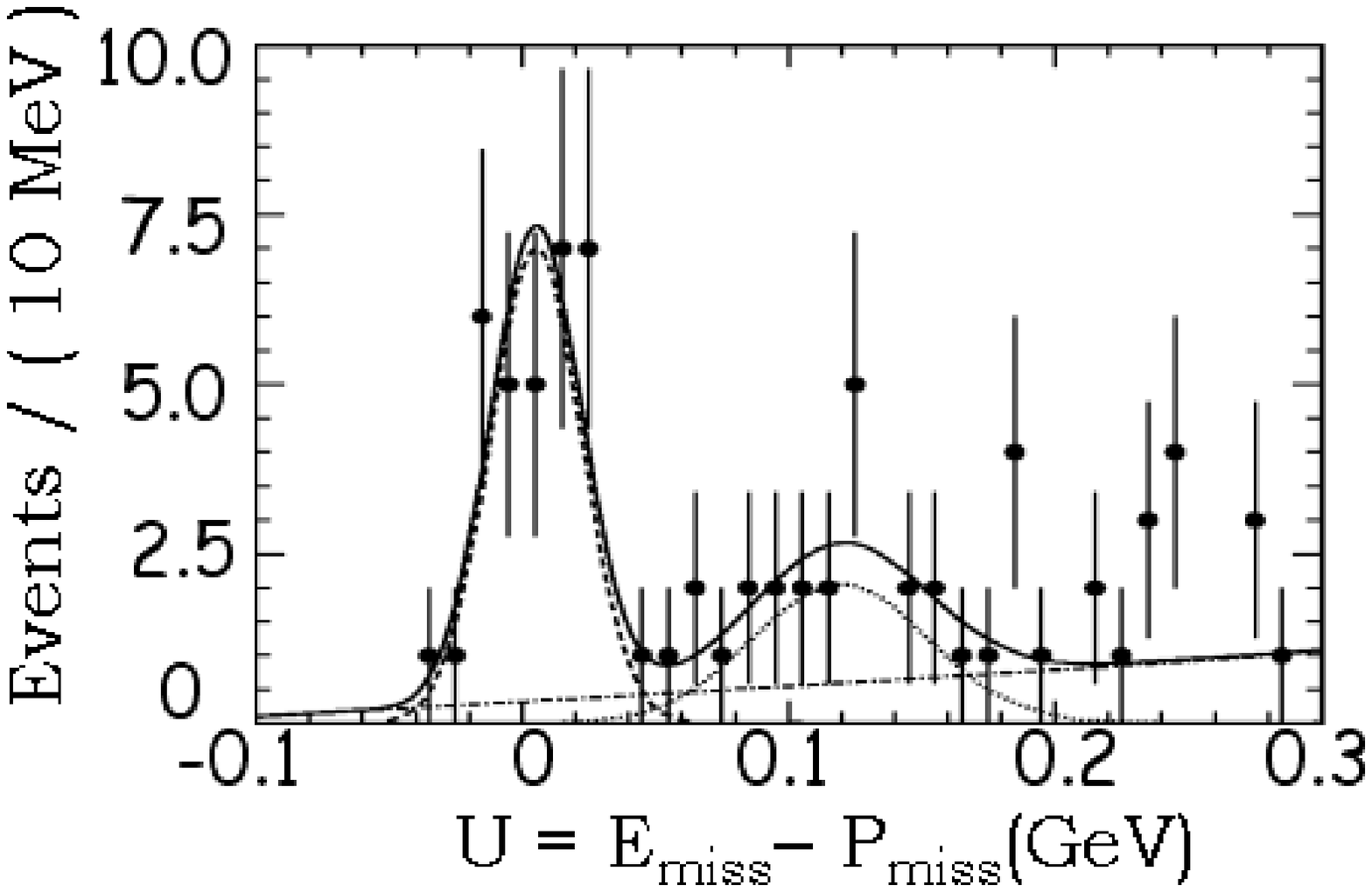}
  }
  \caption{\it Distributions of the variable $U\equiv E_{\rm miss} -
  P_{\rm miss}$ for $D^{0}$ meson decays to a) $\pi^{-}e^{+}\nu$, b)
  $K^{-}e^{+}\nu$, c) $K^{*-}e^{+}\nu$, ($K^{*-}\rightarrow
  K^{-}\pi^{0}$), d) $\rho^{-}e^{+}\nu$.}
  \label{fig:d0exsemi}
\end{figure}

\begin{figure}[htp]
  \centerline{
    {\large a)}
    \epsfxsize=5.7cm
    \epsffile{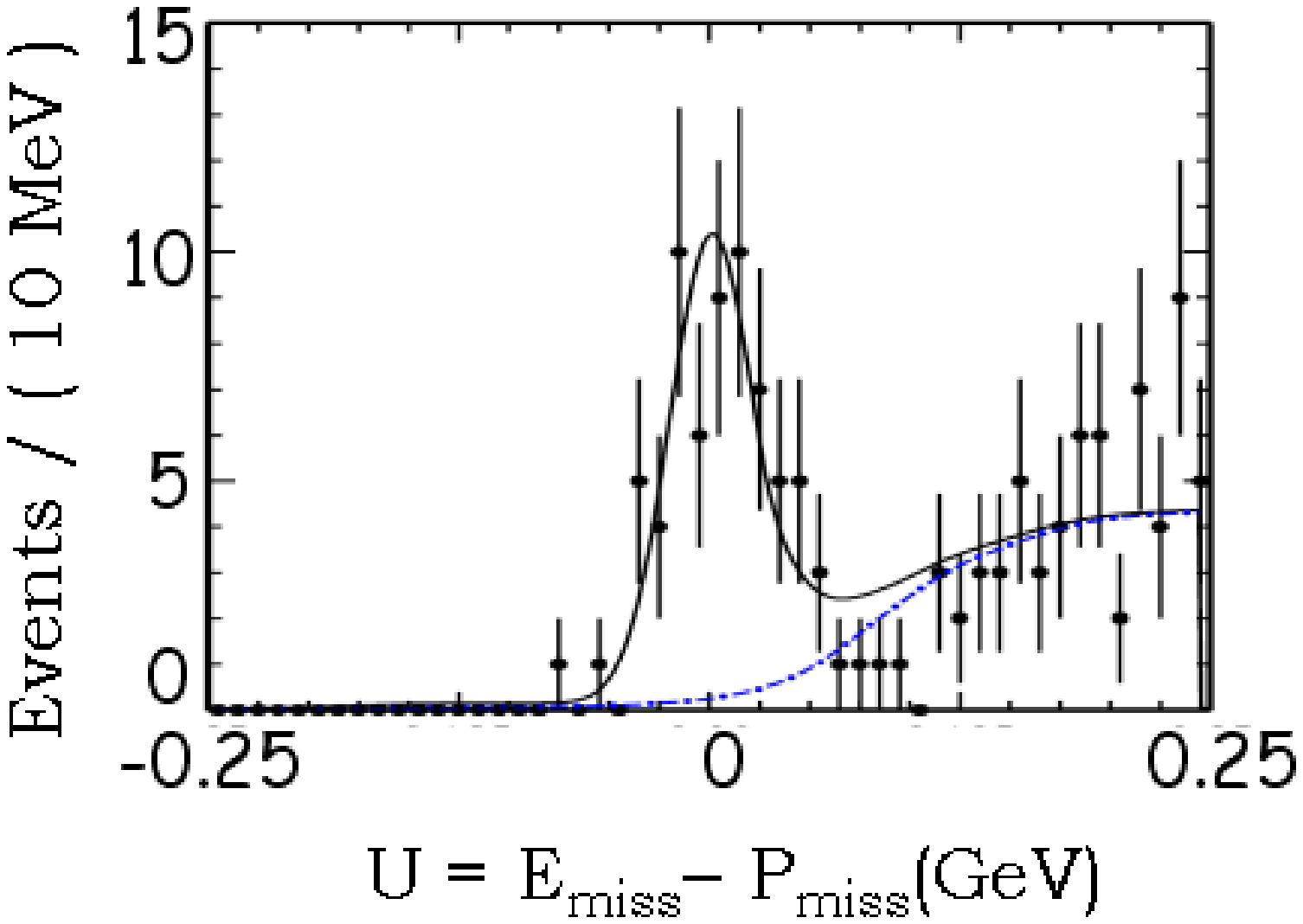}
    {\large b)}
    \epsfxsize=5.7cm
    \epsffile{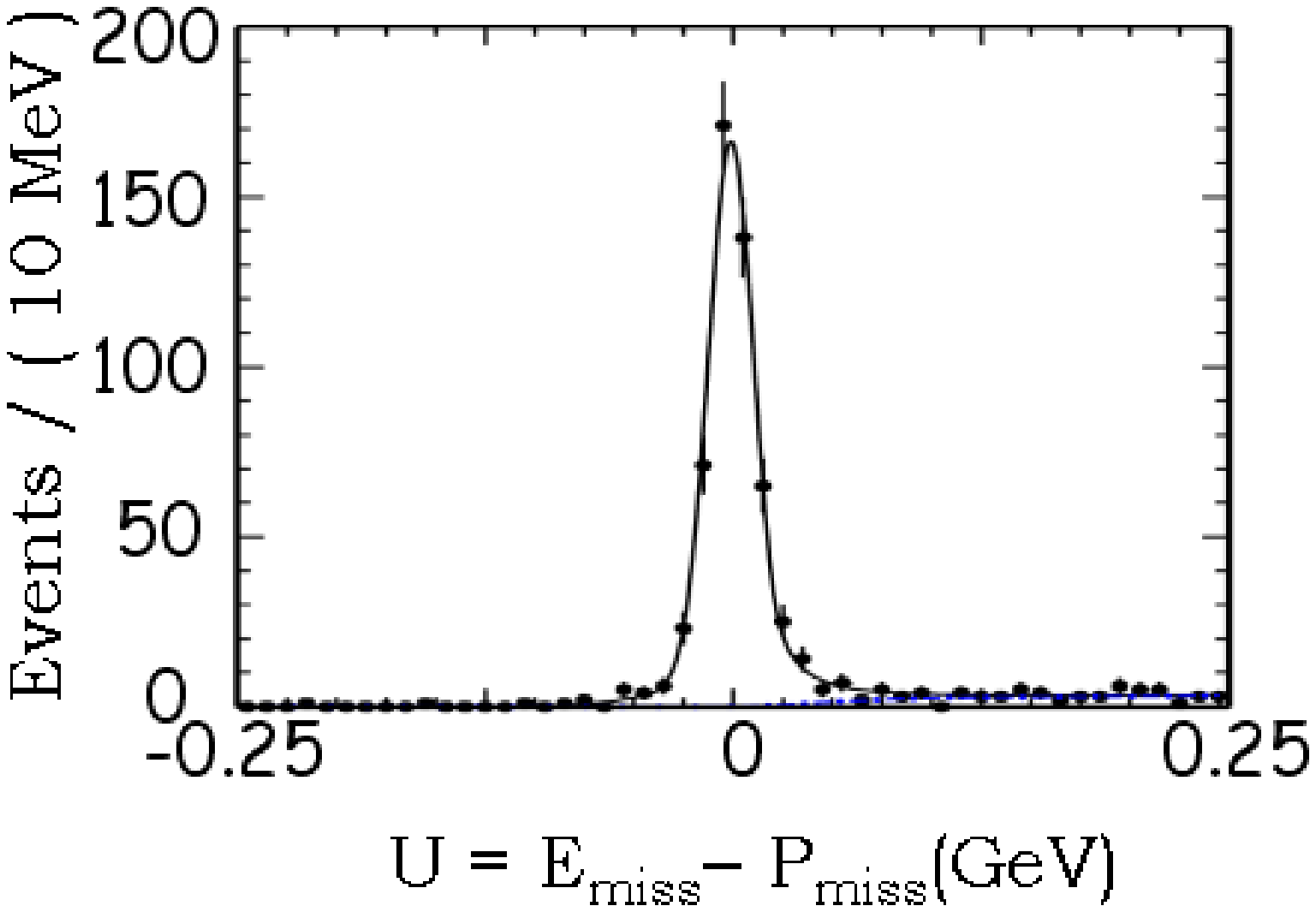}
  }
  \centerline{
    {\large c)}
    \epsfxsize=5.7cm
    \epsffile{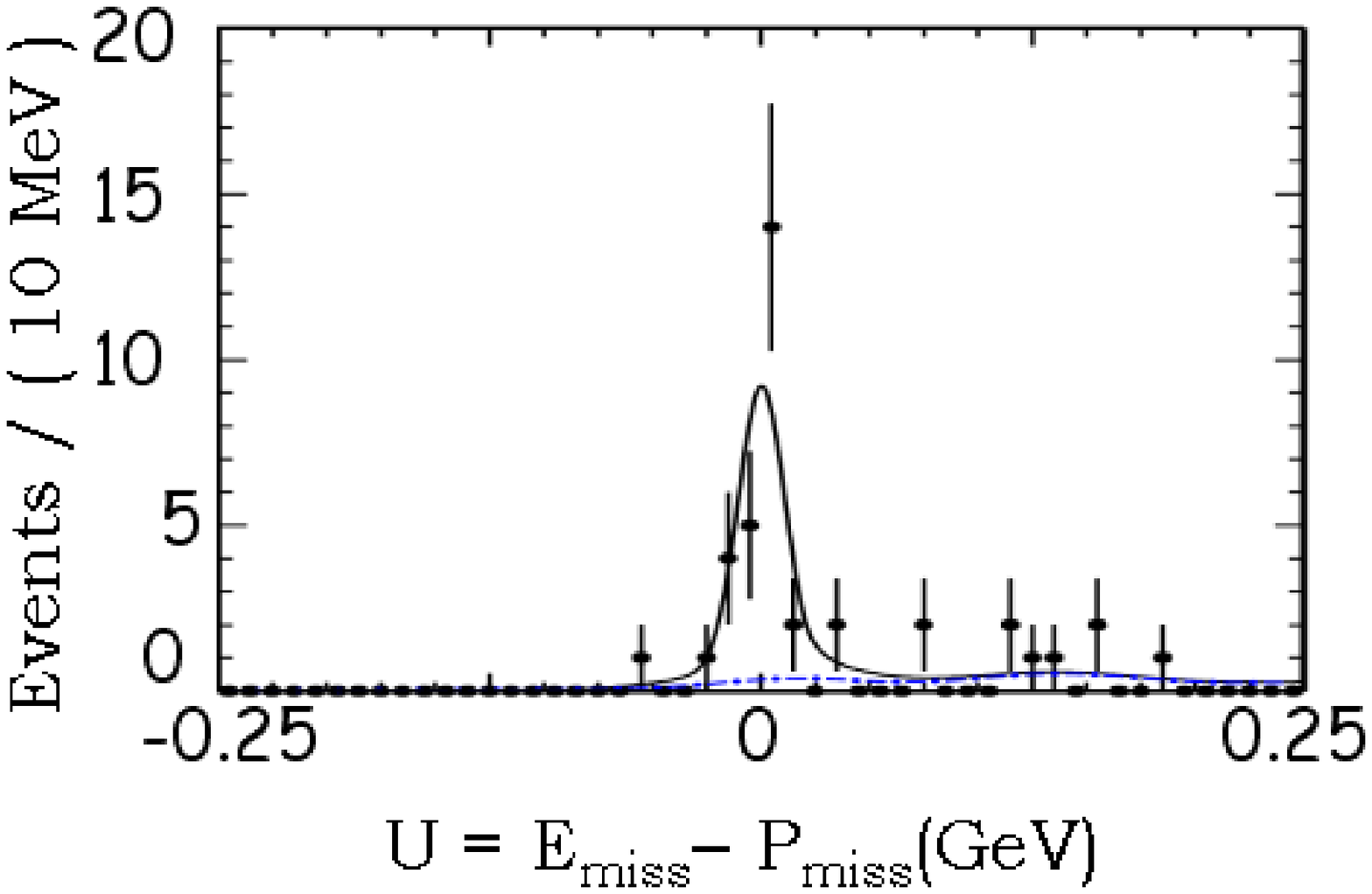}
    {\large d)}
    \epsfxsize=5.7cm
    \epsffile{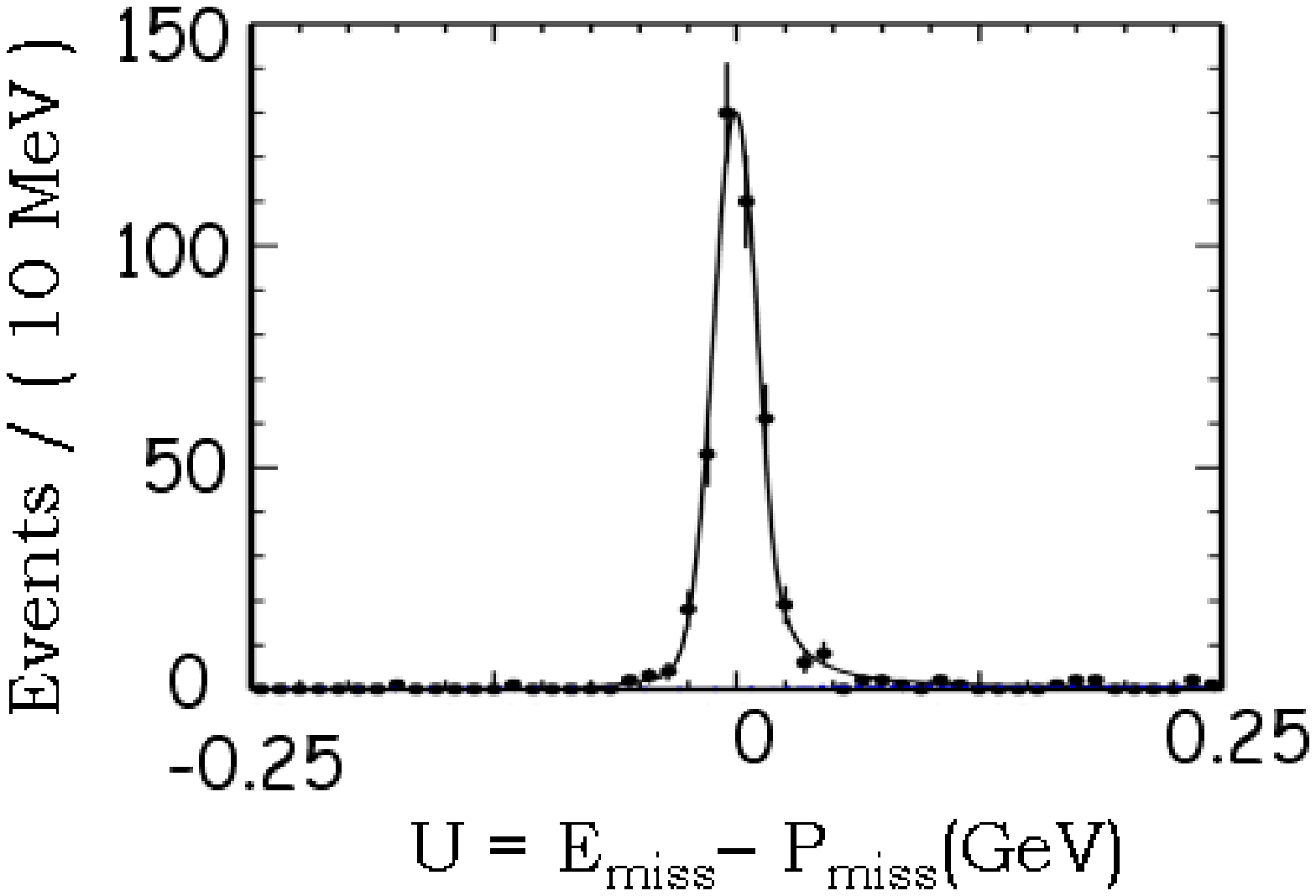}
  }
  \centerline{
    {\large e)}
    \epsfxsize=5.7cm
    \epsffile{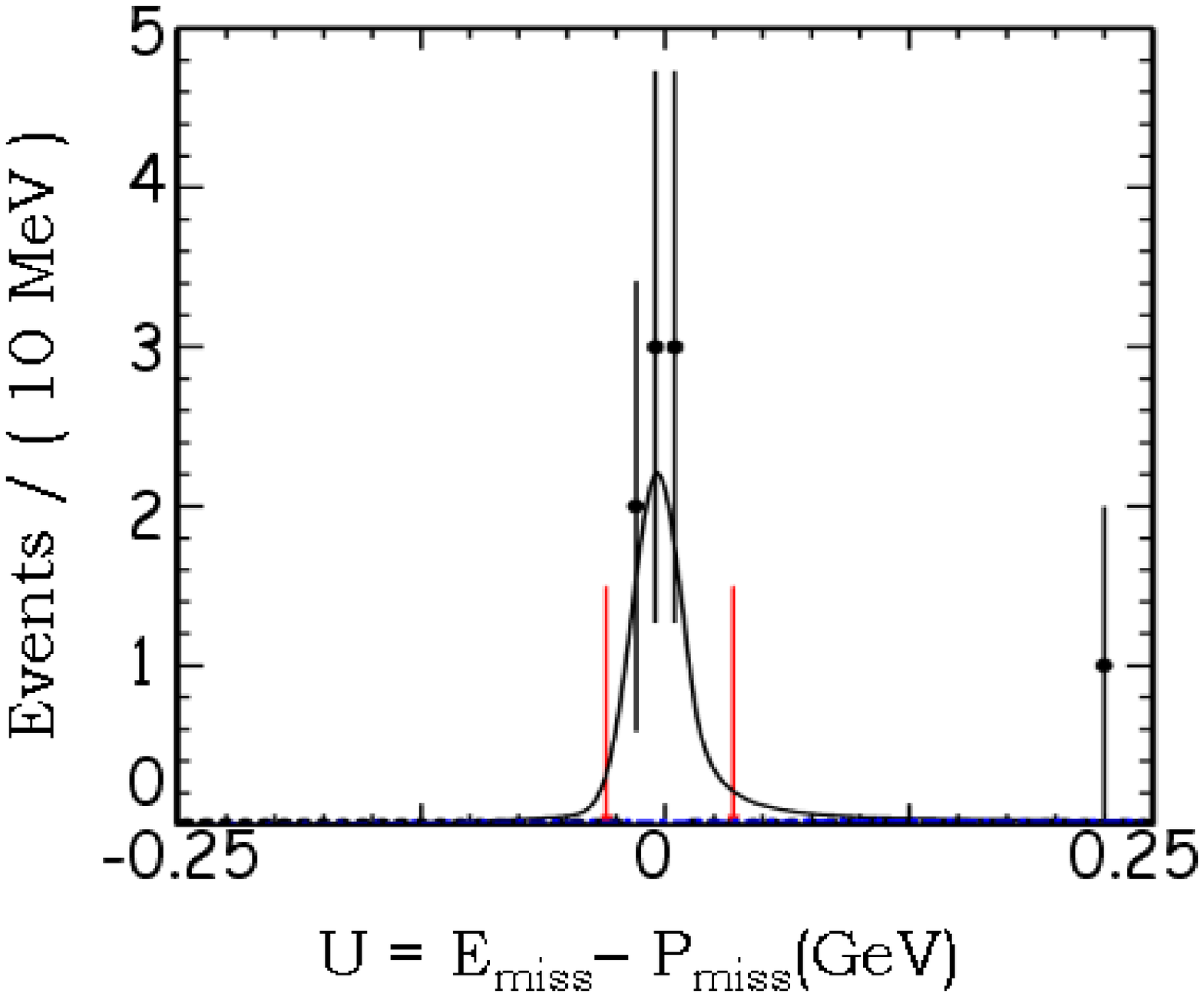}
  }
  \caption{\it Distributions of the variable $U\equiv E_{\rm miss} -
  P_{\rm miss}$ for $D^{+}$ meson decays to a) $\pi^{0}e^{+}\nu$, b)
  $\overline{K^{0}} e^{+} \nu$, c) $\rho^{0} e^{+} \nu$, d)
  $K^{*0}e^{+}\nu$ ($K^{*0}\rightarrow K^{-}\pi^{+}$), e) $\omega
  e^{+}\nu$.} 
  \label{fig:dplusexsemi}
\end{figure}

The analysis of exclusive semileptonic decays also uses a tag of the
other $D$ meson in the event.  These tag $D$ mesons are selected using
the beam-constrained mass of the candidate, defined as 
\begin{equation}
M_{bc} \equiv \sqrt{E_{\rm beam}^{2}  - \vec{p}^{2}_{\rm cand}}, 
\label{eq:Mbc}
\end{equation}
and the energy difference between
the beam and the candidate, defined to be 
\begin{equation}
\Delta E = E_{\rm beam} -E_{\rm cand}.
\label{eq:DeltaE}
\end{equation}
  The remaining observable tracks are then reconstructed
to form the daughter meson.  The missing energy, $E_{\rm miss}$, and missing
momentum, $|\vec{p}_{\rm miss}|$, in the event are used to form the
kinematic variable $U\equiv E_{\rm miss} - |\vec{p}_{\rm miss}|$,
which is fit to determine the signal and background contributions.
These distributions and fits are shown in Fig.~\ref{fig:d0exsemi} and 
Fig.~\ref{fig:dplusexsemi} for the neutral and charged $D$ meson
modes, respectively. 

These raw numbers of events are corrected for efficiency and divided
by the number of tag $D$ mesons to produce absolute branching fractions.
The efficiencies are determined using a combination of GEANT Monte
Carlo and data.  These measurements include first observations of the
modes $D^{0}\rightarrow \rho^{-}e^{+}\nu$ and $D^{+}\rightarrow \omega
e^{+}\nu$.

The absolute branching fraction measurements for these modes are
summarized in Table~\ref{tab:semi}.  Even with only a small fraction
of the final sample, the sensitivity is already an improvement over
previous measurements\cite{PDG} for most modes.

\begin{table}[t]
  \centering
  \caption{ \it Absolute branching fraction measurements by CLEO-c
    (center column) compared with the present measurements tabulated in the
    particle data book\cite{PDG} (right).  All results are preliminary.
    }
  \vskip 0.1 in
  \begin{tabular}{|l|c|c|} \hline
    Decay Mode &  {\cal B} (\%) (CLEO-c) & {\cal B} (\%) (PDG '04) \\
    \hline
    \hline
    $D^{0}\rightarrow \pi^{-}e^{+}\nu$  & $0.25 \pm 0.03 \pm 0.02$  &
    $0.36 \pm 0.06$ \\ 
    $D^{0}\rightarrow K^{-}e^{+}\nu$  & $3.52 \pm 0.10 \pm 0.25$  &
    $3.58 \pm 0.18$ \\ 
    $D^{0}\rightarrow K^{*-}(K^{-}\pi^{0})e^{+}\nu$  & $2.07 \pm 0.23
    \pm 0.18$  &  $2.15 \pm 0.35$ \\ 
    $D^{0}\rightarrow \rho^{-}e^{+}\nu$  & $0.19 \pm 0.04 \pm 0.02$  &
    $-$ \\  \hline
    $D^{+}\rightarrow K^{0}e^{+}\nu$  & $8.71 \pm 0.38 \pm 0.37$  &
    $6.7\pm 0.9$ \\ 
    $D^{+}\rightarrow K^{*0}(K^{-}\pi^{+})e^{+}\nu$  & $5.70 \pm 0.28
    \pm 0.25$  &   $5.5\pm 0.7$ \\ 
    $D^{+}\rightarrow \pi^{0}e^{+}\nu$  & $0.44 \pm  0.06 \pm 0.03$  
     & $0.31\pm 0.15$ \\ 
    $D^{+}\rightarrow \rho^{0}(\pi^{+}\pi^{-})e^{+}\nu$  & $0.21 \pm
    0.04 \pm 0.02$  &   $0.25\pm 0.10$ \\ 
    $D^{+}\rightarrow \omega(\pi^{+}\pi^{-}\pi^{0})e^{+}\nu$  & $0.17 \pm
    0.06 \pm 0.01$  &   $-$ \\ 
    \hline
  \end{tabular}
  \label{tab:semi}
\end{table}

\section{Inclusive Decay Channels}

The inclusive branching fractions ${\cal B}(D^{0}\rightarrow e^{+}X)$ 
and ${\cal B}(D^{+}\rightarrow e^{+}X)$ are being measured by the
CLEO-c collaboration.  At present, the branching fractions are not yet
public, however, even with the preliminary $\sim$60~pb$^{-1}$ sample,
the statistical uncertainties are $\sim 0.2\%$ and $\sim 0.3\%$ for
the $D^{0}\rightarrow e^{+}X$ and $D^{+}\rightarrow e^{+}X$ channels,
respectively.  The corresponding statistical $\oplus$ systematic
uncertainties for the present best measurements are $\sim 0.3\%$ and
$\sim 1.9\%$ for these channels. 

\section{D Meson Absolute Hadronic Branching Fractions and
  $D\overline{D}$ Production Cross Sections}

Using samples of single and double $D$ tagged events,
absolute branching fractions of several hadronic $D$ decay modes are
determined independent of the integrated luminosity, which would
typically add a large uncertainty to the measurement.  This technique
is similar to that used by the Mark III
collaboration\cite{MarkIIIi}\cite{MarkIIIii}.  The number of 
single $D$ tagged events in a given decay mode $i$ is given by 
\begin{equation}
N_{i}= N_{D\overline{D}}{\cal B}_{i}\varepsilon_{i}
\label{eq:Ni}
\end{equation}
 and the number of
double tagged events with decays to modes $i$ and $j$ is given by
\begin{equation}
N_{ij} = N_{D\overline{D}}{\cal B}_{i}{\cal B}_{j}\varepsilon_{ij} .
\label{eq:Nij}
\end{equation}
Equations \ref{eq:Ni} and \ref{eq:Nij} can be combined to give the
number of $D\overline{D}$ pairs produced and the branching fraction of
each mode $i$:
\begin{equation}
N_{D\overline{D}} = \frac{N_{i}N_{j}}{N_{ij}} \frac{\varepsilon_{ij}}
{\varepsilon_{i}\varepsilon_{j}}
\label{eq:NDD}
\end{equation}
and
\begin{equation}
{\cal B}_{i} = \frac{N_ij}{N_{j}}\frac{\varepsilon_{j}}{\varepsilon_{ij}}.
\label{eq:BRi}
\end{equation}

In practice, a simultaneous fit to the neutral modes 
$D^{0}\rightarrow K^{-}\pi^{+}$, $D^{0}\rightarrow
K^{-}\pi^{+}\pi^{0}$, and $D^{0}\rightarrow
K^{-}\pi^{+}\pi^{+}\pi^{-}$ and charged modes $D^{+}\rightarrow
K^{-}\pi^{+}\pi^{+}$ and  $D^{+}\rightarrow K^{0}_{S}\pi^{+}$ is
performed to extract the branching fractions and number of
$D\overline{D}$.  All statistical and systematic correlations between
modes are taken into account in the fit.  The fit is of good quality
with $\chi^{2}/N_{\rm d.o.f} = 9.0/16$ and a confidence level of $91.4\%$.

The efficiencies are
determined from a combination of data and Monte Carlo.  The
denominator of the efficiency calculation may be determined using
missing mass to select events in data and Monte Carlo.  The effects of
final state 
radiation are included in this analysis.

The single and double $D$ tagged yields are determined using the
variables $\Delta E$ and $M_{bc}$, defined in Eqs.~(\ref{eq:DeltaE})
and (\ref{eq:Mbc}), respectively.  Approximately 2500 double tagged
neutral $D$ mesons and 500 double tagged charged $D$ mesons are
reconstructed.

Table~\ref{tab:hadbr} sums up the branching fractions and cross
sections determined from this preliminary analysis.  The statistical
uncertainties on the neutral modes are of order 2.0\% and of order
4.5\% for charged modes.

The uncertainties in the charged
track efficiencies used in this preliminary analysis will be reduced
by about a factor of four in the final analysis of this preliminary
data set.  Improvements to the
$\pi^{0}$ and $K^{0}_{S}$ efficiencies are also nearly complete.  Four
more charged $D$ modes are presently being added and will improve the
statistics in those modes by about a factor of three.  These
measurements will impact the determination of $|V_{cb}|$ by the $B$
factories using $B\rightarrow D^{*} \ell \nu$.

\begin{table}[t]
  \centering
  \caption{ \it Absolute branching fractions and ratios of branching
    fractions of hadronic decays
    and $D\overline{D}$ production cross sections.  All results are
    preliminary. 
    }
  \vskip 0.1 in
  \begin{tabular}{|l|c|} \hline
    Quantity & CLEO-c Measurement \\
    \hline
    \hline
     $\sigma(e^{+}e^{-}\rightarrow D^{0}\overline{D^{0}})$ & $3.47 \pm
    0.07 \pm 0.15$ nb \\ 
     $\sigma(e^{+}e^{-}\rightarrow D^{+}D^{-})$ & $2.59 \pm
    0.11 \pm 0.11$ nb \\ 
     $\sigma(e^{+}e^{-}\rightarrow D\overline{D})$ & $6.06 \pm
    0.13 \pm 0.22$ nb \\ 
     $N_{D^{+}D^{-}}/N_{D^{0}\overline{D^{0}}}$ & $0.75 \pm
    0.04 \pm 0.02$ \\ 
    \hline 
    $N_{D^{0}\overline{D^{0}}}$ & $(1.98 \pm 0.04 \pm 0.03)\times 10^{5}$ \\ 
    ${\cal B}(D^{0}\rightarrow K^{-}\pi^{+})$ & 
    $0.0392 \pm 0.0008 \pm 0.0023$ \\ 
    ${\cal B}(D^{0}\rightarrow K^{-}\pi^{+}\pi^{0})$ & 
    $0.143 \pm 0.003 \pm 0.010$ \\ 
    ${\cal B}(D^{0}\rightarrow K^{-}\pi^{+}\pi^{+}\pi^{-})$ & 
    $0.081 \pm 0.002 \pm 0.009$ \\
    \hline
    $N_{D^{+}D^{-}}$ & $(1.48 \pm 0.06 \pm 0.04)\times 10^{5}$ \\ 
    ${\cal B}(D^{+}\rightarrow K^{-}\pi^{+}\pi^{+})$ & 
    $0.098 \pm 0.004 \pm 0.008$ \\ 
    ${\cal B}(D^{+}\rightarrow K^{0}_{S}\pi^{+})$ & 
    $0.0161 \pm 0.0008 \pm 0.0015$ \\ 
    \hline
    ${\cal B}(D^{0}\rightarrow K^{-}\pi^{+}\pi^{0})/{\cal
    B}(D^{0}\rightarrow K^{-}\pi^{+})$ &   $3.64 \pm 0.05 \pm 0.17$ \\  
    ${\cal B}(D^{0}\rightarrow K^{-}\pi^{+}\pi^{+}\pi^{-})/{\cal
    B}(D^{0}\rightarrow K^{-}\pi^{+})$ &   $2.05 \pm 0.03 \pm 0.14$ \\  
    ${\cal B}(D^{+}\rightarrow K^{0}_{S}\pi^{+})/{\cal
    B}(D^{+}\rightarrow K^{-}\pi^{+}\pi^{+})$ &   $0.164 \pm 0.004 \pm
    0.006$ \\   
    \hline
  \end{tabular}
  \label{tab:hadbr}
\end{table}

\section{Conclusions}
CLEO-c is producing results that will have a large impact on
electroweak physics.  These measurements are essential for the $B$
factories and the Tevatron experiments to realize their full
potential on many measurements.  Using only a small preliminary sample
corresponding to an 
integrated luminosity of 60~pb$^{-1}$ many of these measurements are
already the most significant.  A considerably larger sample is
presently being collected at the $\psi(3770)$. 
The CLEO-c collaboration also plans to
study $D_{s}$ decays in collisions above $D_{s}\overline{D_{s}}$
threshold and study radiative J/$\psi$ decays.

\section{Acknowledgments}
We gratefully acknowledge the effort of the CESR staff
in providing us with excellent luminosity and running conditions.
This work was supported by the National Science Foundation
and the U.S. Department of Energy.


\begin{thebibliography}{99}
\bibitem{CKM} M.~Kobayashi and T.~Maskawa, Prog. Theor. Phys. {\bf
  49}, 652 (1973).
\bibitem{DMuNuPRD} CLEO Collaboration, G.~Bonvicini {\it et al}, Phys. 
  Rev. D {\bf 70}, 112004 (2004).
\bibitem{PDG} Particle Data Group, S.~Eidelman {\it et al}, 
  Phys. Lett. B {\bf 592}, 1 (2004).
\bibitem{MarkIIIi} Mark III Collaboration, R.~M.~Baltrusaitis {\it et
  al}, Phys. Rev. Lett. {\bf 56}, 2140, (1986).
\bibitem{MarkIIIii} Mark III Collaboration, J.~Adler {\it et
  al}, Phys. Rev. Lett. {\bf 60}, 89, (1988).
\end{thebibliography}
\end{document}